%393.tex
%Comm. Math. Phys., 207, N1, (1999), 231-247.
\input amstex

\documentstyle{amsppt}

\hsize 6.25truein
\hfuzz=25pt
\vsize 9.0truein

%\NoBlackBoxes
\TagsOnRight
\leftheadtext{A.G. RAMM}
\rightheadtext{INVERSE SCATTERING}
\define\vl{\varphi_\ell}
\define\iro{\int^r_0}
\define\k{K(r,\rho)}
\define\ul{u_\ell}
\define\dl{\{\delta_\ell\}}
\topmatter
Comm. Math. Phys., 207, N1, (1999), 231-247.
\title   An inverse scattering problem with part of the fixed-energy phase
shifts
  \endtitle
\author  A.G. Ramm  \endauthor
\affil Mathematics Department, Kansas State University,
         Manhattan, KS  66506, USA \\
email:  { ramm\@math.ksu.edu}  
\endaffil
\abstract Assume that $q(r)$ is a real-valued, compactly supported potential,
$q(r)=0$ for $r:=|x|>a$, $q(r)\in L^2(B_a),  \,n=3,\,
B_a:=\{x: x\in \Bbb R^n,\, |x|<a \}$. 
Let $\Cal L$ be an arbitrary
fixed subset of non-negative integers such that $\underset \ell\ne 0,\ell\in
\Cal L\to\sum\frac 1\ell =\infty$, and $\dl$ be fixed-energy phase shifts
corresponding to $q(r)$. The main result is:
\proclaim{Theorem}The data $\dl_{\forall\ell\in\Cal L}$ determine $q(r)$
uniquely.
\endproclaim
\endabstract



\endtopmatter

\vglue 0.2in

\document

\head 1. Introduction  \endhead
In this paper a uniqueness theorem is established for 
inverse scattering
with
fixed energy data. This theorem says that a real-valued
compactly supported spherically
symmetric potential $q(r)$ is uniquely determined by the subset of the phase
shifts $\delta_\ell$ at an arbitrary fixed positive energy.
The subset $\dl_{\ell\in\Cal L}$ 
is defined by an arbitrary subset $\Cal L$ of
integers $\{0,1,2,...\}$, with the property
$$
\underset \ell\in\Cal L,\ell\ne 0\to\sum\frac 1\ell =\infty \tag1.1
$$
No results of this type have been known or
conjectured earlier.

Condition (1.1) appeared in M\"untz's theorem: it is a necessary and
sufficient
condition for the completeness of the set $\{x^\ell\}$ in $L^1(0,a)$, for
an arbitrary fixed $a>0$. 

Such a result gives much deeper understanding of the 
quantum-mechanical inverse
scattering problem 
with fixed-energy data. It may also be of 
some practical significance
because in some physical experiments the phase
shifts can be measured not for all $\ell$ and it is important to
know what part of the fixed-energy phase shifts
is still sufficient for the unique identification of the potential.

We now describe the basic ideas of the proof. 
Our proof is based on two
fundamental results. 

{\it The first result} is the uniqueness theorem from [R1] which
says that the fixed-energy scattering data
 $ A(\alpha^\prime,\alpha),\,\,\,
\forall\alpha^\prime,\alpha\in S^2$, 
determine $q(x)\in Q_a$ uniquely. Here
$A(\alpha^\prime,\alpha)$ is the 
scattering amplitude corresponding to the
potential $q(x)\in Q_a:=\{q:q=\bar q,\,\,q(x)=0$ for $|x|>a,\,\, 
q(x)\in L^2
(B_a)\}$, where 
$B_a:=\{x:|x|\leq a,\,\,x\in\Bbb R^3 \}$ and the
bar stands for complex conjugate.
Under this assumption one can check, using H\"older's
inequality, that $\int_0^a r|q(r)|dr <\infty.$
The proof of the above uniqueness theorem
is valid in $\Bbb R^n,\, n\geq 3$, and
is not valid for $n=2$ (see [R2] for an explanation). 
An algorithm for inversion of noisy fixed-energy
scattering data is developed in [R5] (see also [R2]), where
a stability estimate for this algorithm is obtained.
In [A] a discussion of the Newton-Sabatier
method for inversion of fixed-energy phase shifts
is given and an example is constructed
of two quite different
compactly supported piecewise-constant potentials
which produce practically the same ( within the
accuracy $10^{-5}$) fixed-energy phase
shifts for all values of the angular momenta.
This example illustrates the stability estimate from
[R5] and shows that the above estimate is sharp.
 
Actually, property $C$ for a pair of Schr\"odinger 
operators is used for a proof of the uniqueness theorem
for inverse scattering problem with fixed-energy data.
This theorem follows from property $C$.
The notion of the property $C$ has been introduced and applied to
many inverse problems in a series of papers by the author
 (see [R2] and
references therein, [R3]-[R4]).

Let us formulate this notion for a pair
$\{L_1,L_2\}$ of the Schr\"odinger operators 
$L_j=-\triangledown^2+q_j-k^2,\,\,j=1,2, \,\, q_j\in Q_a,\,\, k\geq 0$
is a constant, $k^2$ is the energy of the particle.

Let $D\subset\Bbb R^n$ be a bounded domain and
$$N_j:=N_D(L_j):=\{w:L_jw=0\,\,\text { in } D,\,\, w \in H^2(D)\}.$$

Let $f\in L^2(D)$.
\definition{Definition 1} If
$$
\{\underset D\to\int f(x)w_1(x)w_2(x)dx=0 \quad\forall w_j\in N_j\}\Rightarrow
f=0, \tag1.2
$$
then we say that the pair $\{L_1,L_2\}$ has property $C$.
\enddefinition
Let $n=3$ and $S^2$ be the unit sphere in $\Bbb R^3$.
 
Define the scattering solution corresponding
to the operator
$L=-\triangledown^2+q-k^2,$ and a fixed $k>0$, which we take,
without loss of generality, to be $k=1$ in this paper, as the solution
to the equation:
$$
[\triangledown^2+1-q(x)]\psi (x,\alpha )=0\text{ in } \Bbb R^3\tag 1.3
$$
$$\psi(x,\alpha)=\exp
(ix\cdot\alpha)+A(\alpha^\prime,\alpha)\frac{e^{ir}}r+
o \left(\frac 1r\right),\quad r:=|x|\rightarrow\infty,\quad \alpha^\prime
:=\frac xr.\tag 1.4
$$
The unit vector $\alpha$ is given, the coefficient
$A(\alpha^\prime,\alpha)$ is
called the scattering amplitude.
It is well known that $q(x)\in Q_a$ determines $A(\alpha^\prime,\alpha)$
uniquely.

It is proved in [R2] that
$$
\{\underset D\to\int f(x)\psi_1(x,\alpha)\overline{\psi_2(x,\beta)}dx=0\;\;
 \forall \alpha,
\beta\in S^2\}\Rightarrow f=0,\tag 1.5
$$
where $\psi_j (x,\alpha),\,j=1,2,$ is the scattering solution
corresponding
to the operator
$L=-\triangledown^2+q_j-k^2, \,\, j=1,2,$ and a fixed $k=1$.

The inverse scattering problem (ISP) with fixed-energy data consists in finding
$q(x)\in Q_a$ given $A(\alpha^\prime,\alpha)\quad 
 \forall\alpha^\prime,\alpha\in S^2$.

Denote by $\tilde S^2$ an arbitrary small open subset of $S^2$.
\proclaim{Theorem 1.1. [R1]} The data $A(\alpha^\prime,\alpha)\quad \forall
\alpha^\prime\in \tilde S^2_1,\quad \forall\alpha\in\tilde S^2_2$, determine
$q(x)\in Q_a$ uniquely.
\endproclaim
\proclaim{Theorem 1.2. [R2]} If $q_j\in Q_a,\quad j=1,2,$ then (1.2)
holds.
\endproclaim
Theorem 1.2 implies that the pair $\{L_1,L_2\}$ of Schr\"odinger's operators
with potentials $q_j\in Q_a$ does have property $C$.

{\it The second result}
we will use is the uniqueness theorem for analytic functions.

 Let us assume that $h(\ell)$ is a holomorphic
function in $\Pi_+:=\{\ell: Re\ell>0\},\quad
\ell=\sigma+i\tau,\quad \sigma \geq 0,$ 
and $\tau$ are real numbers, $h(\ell)\in N$
(Nevanlinna class in $\Pi_+$), that is
$$
\underset 0<r<1\to\sup\int^\pi_{-\pi}\ln^+\biggl | h\left(\frac{1-re^{i\varphi}}
{1+re^{i\varphi}}\right)\biggr |d\varphi<\infty,\tag 1.6
$$
where $\ln^+x= \left\{
\aligned \ln x&\text { if } \ln x>0,\\
              0&\text{ if }\ln x\leq 0.\endaligned
\right.$

\proclaim{Theorem 1.3} If $h(\ell)\in N$ then
$$
h(\ell)=0, \quad\forall\ell\in\Cal L,\tag 1.7
$$
implies
$$
h(\ell)\equiv0\text { in }\Pi_+,\tag1.8
$$
in particular
$$
h(\ell)=0\quad\forall\ell=0,1,2,....\tag1.9
$$
\endproclaim
Theorem 1.3 is a consequence of
Theorem 1.4 which is formulated below. This Theorem
in turn is an immediate corollary to Theorem 15.23 in 
[Ru, p.334].
\proclaim{Theorem 1.4} Assume that the following conditions hold:

i) $f(z)$ is holomorphic in the unit disc $D_1$,

ii) $ f(z)\in N$ in $D_1$, that is
$$
\underset 0<r<1\to\sup\int^\pi_{-\pi}\ln^+|f(re^{i\varphi})|d\varphi<\infty,
\tag 1.10
$$

iii) $f(z_n)=0,\quad n=1,2,3,...,$ 

and

 $$
\sum^\infty_{n=1}(1-|z_n|)=\infty.   \tag 1.11
$$
Then
$$
f(z)\equiv 0   \text { in } D_1.\tag 1.12
$$
\endproclaim
Let us explain why Theorem 1.4 implies Theorem 1.3.
Note that the function
$$
\ell=\frac{1-w}{1+w}\tag1.13
$$
maps conformally $D_1$ onto $\Pi_+$, while
$$
w=\frac{1-\ell}{1+\ell}\tag 1.14
$$
is the inverse map $\Pi_+\rightarrow D_1$.

The function $h(\ell)=h\left(\frac{1-w}{1+w}\right):=
f(w)$ is analytic in $D_1,
\quad f(w)\in N$ in $D_1$, and $f(w_\ell)=0,$ where $w_\ell =\frac{1-\ell}
{1+\ell}$ and $h(\ell)=0$.

If $ \ell\in\Cal L$, then
$$
\underset \ell\in\Cal L,\ell\neq0\to\sum (1-|w_\ell|)=\underset \ell\in\Cal L,
\ell\neq 0\to\sum\left(1-\biggl\vert\frac{1-\ell }{1+\ell}\biggr
\vert
\right)=\underset\ell\in\Cal L,\ell\neq 0\to \sum\frac 2{
\ell+1}=\infty,
\tag1.15
$$
because of the assumption (1.1). Thus, $f(w)=0$ in $D_1$
by Theorem 1.4,  and therefore $h(\ell)=0$ in $\Pi_+$.

Thus,  Theorem 1.3 follows from Theorem 1.4. $\qed$

In order to describe the ideas of our proof, we need to state some known facts
from the scattering theory. If $q(x)=q(r),\, r=|x|$, that is, the potential
is spherically symmetric, and if $k=1$, then the scattering solution is:
$$
\psi (x,\alpha)=\sum^\infty_{\ell =0}4\pi i^\ell\frac{\psi_\ell (r)}r Y_\ell
(x^0)\overline{Y_\ell (\alpha)},\quad x^0:=\frac xr,\tag 1.16
$$
where $\psi_\ell (r)$ satisfies equations (1.17)-(1.19):
$$
\psi^{\prime\prime}_\ell +\psi_\ell -\frac{\ell (\ell +1)}{r^2}\psi_\ell -
q(r)\psi_\ell=0,\quad r>0\tag 1.17
$$
$$
\psi_\ell (r)=O(r^{\ell +1})\text { as } r\rightarrow 0,\tag 1.18
$$
$$
\psi_\ell (r)=e^{i\delta_\ell}\sin (r-\frac{\ell\pi}2 +\delta_\ell)+o(1)
\text { as } r\rightarrow\infty.\tag 1.19
$$
Here $\delta_\ell$ is called the fixed-energy phase shift corresponding to
the angular momentum $\ell$.

The functions $Y_\ell(\alpha)=Y_{\ell m}(\alpha),\quad -\ell\leq m\leq \ell,$
 in (1.12) are the spherical harmonics
orthonormalized in $L^2(S^2)$.The summation in (1.16) and
in (1.20) below
includes the summation with respect to $m, \quad -\ell\leq m\leq\ell$, and is
not shown for brevity.

The corresponding scattering amplitude for the spherically symmetric potential
is of the form:
$$
A(\alpha^\prime,\alpha)=A(\alpha^\prime\cdot\alpha)=\sum^\infty_{\ell=0}
A_\ell Y_\ell (\alpha^\prime)\overline{Y_\ell(\alpha)}, \tag 1.20
$$
where
$$
A_\ell =2\pi i(1-e^{2i\delta_\ell})=4\pi e^{i\delta_\ell}\sin\delta_\ell.
\tag 1.21
$$
Recall that we assume $k=1$ throughout.

Therefore, in the case of spherically symmetric potentials which
we consider in this paper,
there is a one-to-one correspondence between the scattering
amplitude $A(\alpha^\prime,\alpha)$ and the 
set of numbers $\{A_\ell\}_{\ell=0,1,2,...}.$ The
 set $\{\delta_\ell\}_{\ell=0,1,2,...},\quad -\pi\leq\delta_\ell<\pi$,
determines the set $\{A_\ell\}_{\ell =0,1,2,...}$ uniquely,
so there is  a one-to-one correspondence between the scattering amplitude
at a fixed energy and the set of all phase shifts with this
energy.

Using (1.16) and the orthonormality of the spherical harmonics, one obtains:
$$
\{0=\int_{B_a}p(r)\psi_1(x,\alpha)\overline{\psi_2(x,\beta)}dx\} 
 \Longleftrightarrow 
\{0=\sum^\infty_{\ell=0}\int^a_0p(r)\psi_{1\ell}(r)\overline{\psi_{2\ell}(r)}
dr\,\, \overline{Y_\ell(\alpha)}Y_\ell(\beta),\quad \forall\alpha,\beta\in
S^2\},
\tag 1.22
$$
where $p(r)\in Q_a$ is an arbitrary function.

Multiplying the second integral in
(1.22) by $Y_\ell(\alpha)$,  integrating over $S^2$
and using the orthonormality of the spherical harmonics, one
gets that the first equality in (1.22) is equivalent to
$$
0=\int^a_0drp(r)\psi_{1\ell}(r)\overline{\psi_{2\ell}(r)}\quad\quad
\forall\ell=0,1,2,....\tag 1.23
$$
The regular solution $\varphi_\ell (r)$ to equation (1.17) is defined uniquely
by its behavior near the origin:
$$
\vl (r)=\frac{r^{\ell +1}}{(2\ell +1)!!}+o (r^{\ell +1}),
\quad\quad r\rightarrow 0.
\tag 1.24
$$
This solution is a real-valued function for $\ell=0,1,2,...$ and $r>0$. Its
behavior at infinity is known:
$$
\vl =|F_\ell|\sin\left(r-\frac{\ell\pi}2 +\delta_\ell\right)+o(1),\quad
r\rightarrow +\infty,\tag 1.25
$$
where $\delta_\ell$ is the same as in (1.19) and $|F_\ell|$ is a certain
positive constant (the value of the Jost function $F_\ell(k)$ at $k=1$). 
Since $\psi_\ell (r)$ solves (1.17) and satisfies (1.18),
it follows that
$$
\psi_{\ell} =c_\ell\vl (r),\quad\quad c_\ell =const.\tag 1.26
$$
Therefore condition (1.23) is equivalent to
$$
0=\int^a_0drp(r)\varphi_{1\ell}(r)\varphi_{2\ell}(r):=h(\ell),\quad
\ell =0,1,2,...,\tag 1.27
$$
where we have used the real-valuedness of $\vl (r)$.

Let us now describe {\it the idea of our proof}.
\subhead Step 1\endsubhead Assuming that the data $\dl_{\ell\in\Cal L}$
correspond to two different potentials $q_1(r)$ and $q_2(r),\quad q_j(r)\in
Q_a$, we derive the following orthogonality relation:
$$
h(\ell)=0\quad\quad \forall\ell\in\Cal L,\quad \quad p(r):=q_1(r)-q_2(r).
\tag 1.28
$$
\subhead Step 2\endsubhead
We prove that the function $h_1(\ell)$, defined
below, in formula (3.65'), is holomorphic 
in $\Pi_+:=\{\ell:\ell\in
\Bbb C, \Re \ell>0\}$ and belongs to class $N$ defined in (1.6).

Condition (1.28) implies $h_1(\ell)=0 \quad \forall\ell\in\Cal L$.
This implies by Theorem 1.3 that $h_1(\ell)=0 \quad \forall\ell\in \Pi_+$.
Therefore, condition (1.28) implies $h(\ell)=0,\quad\ell=
0,1,2,...,$ that is (1.27) holds.  

This implies, as we have proved above, that
$$
0=\int_{B_a}p(r)\psi_1(x,\alpha)\overline{\psi_2(x,\beta)}dx\quad\quad
\forall\alpha,\beta\in S^2.\tag 1.29
$$
Equation (1.29) and property $C$ for the pair $\{L_1,L_2\}$ of the
Schr\"odinger operators $L_j=-\triangledown^2+q_j(x)
-1,\quad q_j(x)\in Q_a$,
imply $p(r)=0$, that is, $q_1(r)=q_2(r)$.   $\square$

An essential ingredient of our proof of the  implication
$$\{h_1(\ell)=0 \quad \forall\ell\in\Cal L\}\Rightarrow
\{h_1(\ell)=0 \quad\forall\ell=0,1,2,3,....\}.$$
The proof of this implication is based on the existence of 
the transformation operators whose kernel does not
depend on $\ell$. Existence and uniqueness of such operators
as well as the estimate  (3.53) (see section 3 below),
which we use in the proof of the above implication,
are established in section 3. These results, although new
and of independent interest, play an auxiliary role in our proof. 
They are presented in sections 3.3 and 3.4 as a
part of the proof.
 
{\it The description of the idea of our proof is complete.}

In section 3 we derive the orthogonality relation (1.28). 

In the same section we study the analytic properties of $h(\ell)$ as a
function of complex $\ell$. In this study there are two basic steps.

First, we study the function
$$
h_0(\ell):=\int^a_0drp(r)u^2_\ell(r),\quad\quad u_\ell(r):=\sqrt{\frac{\pi r}2}
J_{\ell+\frac 12}(r),\tag 1.30
$$
where $J_{\ell+\frac 12}(r)$ is the standard Bessel function.

Note that $\vl (r)=u_\ell(r)$ if $q(r)=0$.

Define
$$
H(\ell):=
h_0(\ell)[\sqrt{\frac 2\pi}\Gamma\left(\frac 12\right)2^{\ell+\frac
12}\Gamma (\ell +1)]^2,\tag 1.31
$$
where $\Gamma (z)$ is the Gamma-function.

We prove:
\proclaim {Lemma 1.1} The function $H(\ell)$ is holomorphic
in $\Pi_+$ and
$H(\ell)\in N$\endproclaim
Secondly, we prove the existence of the transformation operator, which sends
$\ul (r)$ into $\vl (r)$:
$$
\vl (r)=\ul (r)+\iro\k\ul (\rho)\rho^{-2}d\rho,\quad K(r,0)=0.\tag 1.32
$$
It is crucial for our argument that $\k$ does not depend on $\ell$.
Therefore, the analytic properties of 
$h_1(\ell):=h(\ell)[\Gamma\left(\frac 12\right)
\Gamma \left(
\ell+\frac 12\right)2^{\ell+\frac 12}\sqrt{\frac 2\pi}]^2$ 
and $H(\ell)$, as functions of
$\ell$, are essentially the same: these two functions 
are both holomorphic in $\Pi_+$ and belong
to class $N$.

In section 3 we prove some technical estimates for the kernel $\k$ of the
transformation operator.

Although the transformation operators of the type (1.32) appeared formally
earlier in the physical literature [CS, p.185], their existence was not proved.
In the literature there exists a construction of the transformation
operators whose kernels depend on $\ell$, see [V], [Le], [M].

The difficulty of the existence proof
for the transformation operator, whose kernel $\k$ does not depend
on $\ell$, comes from the fact that the Goursat-type
problem which one can derive for $\k$ involves 
differential operators with variable
coefficients which degenerate at the origin. 

We overcome this difficulty by introducing new variables and reducing the
problem to an equivalent Volterra-type integral equation.

Existence and uniqueness of the solution to this equation are established in
section 3, where some estimates of the solution are given.

This concludes the introduction.

In section 2 we state the basic uniqueness result.

In section 3 proofs are given.
\head 2. Statement of the basic result\endhead
Let us assume that
$$
q(x)\in Q:=\{q: q\in Q_a, \quad q(x)=q(r),\,r:=|x|\},\tag 2.1
$$
and let $\delta_\ell$ denote the fixed-energy phase shifts. Note that
if $q\in Q$, then $\int_0^ar|q(r)|dr<c<\infty.$ Here and below $c>0$
stands for various estimation constants.

The inverse scattering problem we are interested in can now be formulated:

ISP. {\it Given the data $\dl_{\forall\ell\in\Cal L}$, where $\Cal L$
satisfies condition (1.1), can one recover $q(r)\in Q$ uniquely?}

Our basic result is:
\proclaim {Theorem 2.1} Let $\Cal L$ be an arbitrary fixed subset of
non-negative integers which satisfies condition (1.1). Then the data
$\dl_{\forall\ell\in\Cal L}$, corresponding to a $q(r)\in Q$, 
determine $q(r)$ uniquely.
\endproclaim
This result implies, in particular, that there is no $q(r)\in Q,\quad
q(r)\not\equiv 0$, such that $\delta_{2\ell}=0\quad \forall\ell =0,1,2,...$.

It also implies that there is no $q(r)\in Q$, such that $\delta_0\neq
0,\, \delta_\ell=0, \,\, \ell=1,2,3,....$, which means that
there are no potentials in $Q$ producing the scattering
amplitude $A(\alpha, \alpha')$
which is constant for all $\alpha, \alpha' \in S^2$ at a
fixed positive energy, see also [R6] where this was
proved for the first time by a different argument. 

\head 3.Proofs\endhead
\subhead 3.1. Proof of the orthogonality relation (1.28)\endsubhead

Suppose $q_1(r)$ and $q_2(r)$ generate 
the same data $\dl_{\forall\ell\in\Cal L}$.
Subtract from equation (1.17) with $q=q_1$ and $\psi_\ell =\psi_{1\ell}$
similar equation with $q=q_2$ and $\psi_\ell =\psi_{2\ell}$ to get:
$$
\psi^{\prime\prime}_\ell +\psi_\ell -\frac{\ell(\ell+1)}{r^2}\psi_\ell-q_1
\psi_\ell =p(r)\psi_{2\ell},\tag 3.1
$$
where
$$
\psi_\ell:=\psi_{1\ell}-\psi_{2\ell},\quad\quad p(r):=q_1(r)-q_2(r).\tag 3.2
$$
Multiply (3.1) by $\psi_{1\ell}(r)$, integrate over $(0,\infty)$ and then by
parts using (1.18) and (1.19) and the assumption that $\delta_\ell$ is the same
for $\psi_{1\ell}$ and $\psi_{2\ell}$ for $\ell\in \Cal L$. The result is
$$
0=\int^a_0p(r)\psi_{1\ell}(r)\psi_{2\ell}(r)dr\quad\quad \forall\ell\in\Cal L.
\tag3.3
$$
This is equivalent to the desired relation (1.28) because of (1.26). $\square$
\subhead 3.2. Analytic properties of the function $H(\ell)$\endsubhead
\demo {Proof of Lemma 1.1} Recall the well-known formula [GR, 8.411.8]:
$$
\sqrt{\frac 2\pi}\Gamma\left(\frac 12\right)2^{\ell+\frac 12}\Gamma (\ell +1)
u_{\ell}(r)
=r^{\ell + 1}\int^1_{-1}(1-t^2)^\ell e^{irt}dt.\tag 3.4
$$
From (1.30), (1.31), and (3.4) one gets:
$$
H(\ell)=\int^a_0drp(r)r^{2\ell +2}\left(\int^1_{-1}(1-t^2)^\ell e^{irt}dt\right)
^2.\tag3.5
$$
Let $\ell =\sigma +i\tau,\quad \sigma\geq 0$. Then $H(\ell)$ is a 
holomorphic
function of $\ell$ for $\sigma >0$ and
$$
|H(\ell)|\leq\int^a_0dr|p(r)|r a^{ 2\sigma+1 } \leq ca^{ 2\sigma}.
\tag3.6
$$
One can always assume $a>1$ since $\sigma>0$.

Let us check that (3.6) implies that $H(\ell)\in N$. One has
$$\ln^+(ab)\leq
\ln^+a+\ln^+b \quad \text { for } a,b>0.
$$
Therefore, using (3.6), one obtains:
$$
\multline
\int^\pi_{-\pi}\ln^+\vert
H\left(\frac{1-re^{i\varphi}}{1+re^{i\varphi}}\right)
\vert d\varphi\leq\int^\pi_{-\pi}\ln^+\vert ca^{2Re\frac{1-re^{i\varphi}}{1+
re^{i\varphi}}}\vert d\varphi\leq \\
\leq c_1+2\ln a\int^\pi_{-\pi}Re\frac
{1-re^{i\varphi}}{1+re^{i\varphi}}d\varphi\leq c_1+2\ln a\int^\pi_{-\pi}\frac
{1-r^2}{1+r^2+2r\cos\varphi}d\varphi =c_1+4\pi\ln a<\infty, \, a>1.
\endmultline\tag3.7
$$
Here we have used the known formula
$$
\int^\pi_{-\pi}\frac{d\varphi}{1+r^2+2r\cos\varphi}=\frac{2\pi}{1-r^2},\quad
\quad 0<r<1,\tag3.8
$$
which is easy to check.

Estimate (3.7) proves that $H(\ell)\in N$. Lemma 1.1 is proved. $\square$
\enddemo
\subhead 3.3 Transformation operators\endsubhead

Define
$$
L_r\varphi :=\left[r^2\frac{\partial^2}{\partial r^2}+r^2-r^2q(r)\right]\varphi:=L_{0r}
\varphi -r^2q(r)\varphi.\tag3.9
$$
For the regular solution to (1.17)
one has the following differential equation:
$$
L_r\varphi_\ell (r)=\ell(\ell +1)\varphi_\ell(r),\tag3.10
$$
and for the function
$\ul (r) =\sqrt{\frac{\pi r}2} J_{\ell+\frac 12}(r)$ the equation
$$
L_{0r}\ul (r)=\ell(\ell+1)\ul (r).\tag3.11
$$
Let us look for the kernel $\k$ such that formula (1.32) gives the regular
solution to equation (1.17). Substitute (1.32) into (1.17), drop index $\ell$
for convenience, use (3.10) and (3.11), and get
$$
\multline
0=-r^2q(r)u+(r^2-r^2q(r))\iro\k u\rho^{-2}d\rho\\
-\iro\k \rho^{-2} L_{0\rho}u d\rho +
r^2\partial^2_r\iro\k u\rho^{-2}d\rho.
\endmultline\tag3.12
$$
We assume first
that $\k$ is twice continuously differentiable with respect to its
variables in the region $0<r<\infty,\quad 0<\rho\leq r$. This 
assumption requires
extra smoothness of $q(r),\quad q(r)\in C^1(0,a)$. If $q(r)$ satisfies condition
(2.1), then equation (3.18) below has to be understood in the sense of
distributions. Eventually we will work with an integral equation (3.45)
(see below) for which assumption (2.1) suffices.

Note that
$$
\iro\k\rho^{-2}L_{0\rho}ud\rho=
\iro L_{0\rho}\k u\rho^{-2}d\rho+K(r,r)u_r-K_\rho (r,r)u,
\tag3.13
$$
provided that
$$
K(r,0)=0. \tag3.14
$$
We assume (3.14) to be valid.

Denote
$$
\dot K:=\frac {dK(r,r)}{dr}.\tag3.15
$$
Then
$$
\multline
r^2\partial^2_r\iro \k u\rho^{-2}d\rho=\dot Ku+K(r,r)u_r-\frac 2r K(r,r)u+\\
K_r(r,r)u+r^2\iro K_{rr}(r,\rho)u\rho^{-2}d\rho.
\endmultline\tag3.16
$$

Combining (3.12)-(3.16) and writing again $\ul$ in place of $u$, one gets
$$\multline
0=\iro [L_r\k -L_{0\rho}\k]\ul(\rho)\rho^{-2}d\rho+\ul(r)[-r^2q(r)+
\dot K-\\
\frac{2K_r(r,r)}r+K_r(r,r)+K_\rho(r,r)],\quad \forall r>0,
\quad \ell=0,1,2,....\endmultline\tag3.17
$$
Let us prove that (3.17) implies:
$$
L_r\k =L_{0\rho}\k,\quad 0<\rho\leq r,\tag3.18
$$
$$
q(r)=\frac{2\dot K}{r^2}-
\frac{2K(r,r)}r=\frac 2r\frac d{dr}\frac{K(r,r)}r.\tag3.19
$$
This proof requires a lemma.
\proclaim {Lemma 3.1} Assume that $\rho f(\rho)\in L^1(0,r)$
and $\rho A(\rho)\in L^1(0,r)$.  If
$$
0=\iro f(\rho)\ul (\rho)d\rho+\ul (r)A(r) \quad \forall 
\ell=0,1,2,...,\tag3.20
$$
then
$$
f(\rho)\equiv 0\text{ and } A(r)=0.\tag3.21
$$
\endproclaim
\demo{Proof} Equations (3.20) and (3.4) imply:
$$
\gather
0=\int^1_{-1}dt(1-t^2)^\ell\left(\frac d{idt}\right)^\ell\iro d\rho\rho f(\rho)e^
{i\rho t}+\\
rA(r)\int^1_{-1}(1-t^2)^\ell\left(\frac d{idt}\right)^\ell e^{irt}dt
\endgather
$$
Therefore
$$
0=\int^1_{-1}dt\frac{d^\ell (t^2-1)^\ell}{dt^\ell}
[\iro d\rho\rho f(\rho)e^
{i\rho t}+rA(r)e^{irt}],\quad l=0,1,2,....\tag 3.22
$$
Recall that the Legendre polynomials are defined by the formula
$$
P_\ell (t)=\frac 1{2^\ell !}\frac{d^\ell}{dt^\ell}(t^2-1)^\ell\tag3.23
$$
and they form a complete system in $L^2(-1,1)$.

Therefore (3.22) implies
$$
\iro d\rho\rho f(\rho)e^{i\rho t}+rA(r)e^{irt}=0\quad \forall t\in [-1,1].
\tag3.24
$$
Equation (3.24)implies
$$
\iro d\rho\rho f(\rho)e^{i\rho t}=0,\quad \forall t\in [-1,1],\tag3.25
$$
and
$$
rA(r)=0.\tag3.26
$$
Therefore $A(r)=0$. Also $f(\rho)=0$ because the left-hand side of (3,25) is an
entire function of $t$, which vanishes on the interval $[-1,1]$ and,
consequently, it vanishes identically, so that $\rho f(\rho) =0$ and therefore
$f(\rho)\equiv 0$.

Lemma 3.1 is proved.$\square$
\enddemo
\subhead 3.4. Existence and uniqueness 
of the transformation operators\endsubhead

Let us prove that the problem (3.18), (3.19), (3.14), which is a Goursat-type
problem, has a solution and this solution is unique in the class of functions
$\k$, which are twice continuously differentiable with respect to $\rho$ and
$r,\quad 0<r<\infty,\quad 0<\rho\leq r$.
In this section we assume that $q(r)\in C^1(0,a)$. This assumption implies that
$\k$ is twice continuously differentialable. If $q(r)\in Q$, see (2.1), 
the
arguments in this section which deal with integral equation (3.45) remain valid.
Specifically, existence and uniqueness of the solution to 
equation (3.45) is proved under the only assumption $\int_0^a r|q(r)|dr<\infty$
as far as the smoothness of $q(r)$ is concerned.

By a limiting argument one can reduce the smoothness requirements on $q$ to the
condition (2.1) but in this case equation (3.18) has to be understood in
distributional sense.

Let us rewrite the problem we want to study:
$$
r^2K_{rr}-\rho^2 K_{\rho\rho}+
[r^2-r^2q(r)-\rho^2]\k=0,\quad 0<\rho\leq r,\tag3.27
$$
$$
K(r,r)=\frac r2\iro sq(s)ds:=g(r),\tag3.28
$$
$$
K(r,0)=0.\tag3.29
$$
The difficulty in the study of this Goursat-type problem comes from the fact
that the coefficients in front of the second derivatives of the kernel
$\k$
are variable.

Let us reduce problem (3.27)-(3.29) to the one with constant coefficients. To
do this, introduce the new variables:
$$
\xi=\ln r+\ln\rho,\quad \eta=\ln r-\ln\rho. \tag3.30
$$
Note that
$$
r=e^{\frac{\xi +\eta}2},\quad \rho =e^{\frac{\xi-\eta}2},\tag3.31
$$
$$
\eta\geq0,\quad -\infty<\xi<\infty,\tag3.32
$$
and
$$
\partial_r=\frac 1r(\partial_\xi +\partial_\eta),\quad \partial_\rho =
\frac 1\rho (\partial_\xi -\partial_\eta).\tag3.33
$$
Let 
$$\k :=B(\xi,\eta).$$ 
A routine calculation transforms equations (3.27)-(3.29)
to the following ones:
$$
B_{\xi \eta}(\xi,\eta)-\frac 12 B_\eta (\xi,\eta) +Q(\xi,\eta)B=0,
\quad \eta\geq0,\quad
-\infty<\xi<\infty,\tag3.34
$$
$$
B(\xi,0)=g\left(e^{\frac \xi 2}\right):=G(\xi),\quad -\infty<\xi<\infty\tag3.35
$$
$$
B(-\infty,\eta)=0,\quad \eta\geq0,\tag3.36
$$
where $g(r)$ is defined in (3.28).

Here we have defined
$$
Q(\xi,\eta):=\frac 14\left[e^{\xi +\eta} -e^{\xi +\eta}q\left(e^{\frac{\xi+
\eta}2}\right)-e^{\xi-\eta}\right],\tag3.37
$$
and took into account that $\rho=r$ implies $\eta =0$, while $\rho =0$ implies,
for any fixed $\eta\geq 0$, that $\xi=-\infty$.

Note that
$$
 \underset -\infty<\xi<\infty
\to\sup e^{-\frac \xi 2}G(\xi)<c,\tag3.38
$$
$$
\underset 0\leq\eta\leq B\to \sup \int^A_{-\infty}|Q(s,\eta)|ds\leq c(A,B),
\tag3.39
$$
for any $A\in\Bbb R$ and $B>0$, where $c(A,B)>0$ is a constant.

To get rid of the second term on the left-hand side of (3.34), let us introduce
the new kernel $L(\xi,\eta)$ by the formula:
$$
L(\xi,\eta):=B(\xi,\eta)e^{-\frac \xi 2}.\tag3.40
$$
Then (3.34)-(3.36) can be written as:
$$
L_{\eta \xi}(\xi,\eta )+Q(\xi,\eta)L(\xi,\eta)=0,\quad \eta\geq 0,\quad
-\infty<\xi<\infty, \tag3.41
$$
$$
L(\xi,0)=e^{-\frac\xi 2}G(\xi):=b(\xi):=
\frac 12\int^{e^{\frac \xi 2}}_0 s q(s) ds,
\quad -\infty<\xi<\infty, \tag3.42
$$
$$
L(-\infty,\eta)=0,\quad \eta\geq 0.\tag3.43
$$
We want to prove existence and uniqueness of the solution to (3.41)-(3.43).
In order to choose a convenient Banach space in which to work, let us transform
problem (3.41)-(3.43) to an equivalent Volterra-type integral equation.

Integrate (3.41) with respect to $\eta$ from $0$ to $\eta$ and use (3.42) to get
$$
L_\xi (\xi,\eta)-b^\prime (\xi)+\int^\eta_0 Q(\xi,t)L(\xi, t)dt=0.\tag3.44
$$
Integrate (3.44) with respect to $\xi$ from $-\infty$ to $\xi$ and use (3.44) to
get
$$
L(\xi,\eta
)=-\int^\xi_{-\infty}ds\int^\eta_0dtQ(s,t)L(s,t)+b(\xi):=VL+b,\tag3.45
$$
where
$$
VL:=-\int^\xi _{-\infty}ds\int^\eta_0dtQ(s,t)L(s,t).\tag3.46
$$
Consider the space $X$ of continuous functions $L(\xi,\eta)$, defined in the
half-plane $\eta\geq 0,\quad -\infty<\xi<\infty$, such that for any $B>0$ and
any $-\infty<A<\infty$ one has
$$
\Vert L\Vert :=\Vert L\Vert_{AB}:=\underset -\infty<s\leq A\to{\underset
0\leq t\leq B\to\sup}\left(e^{-\gamma t}|L(s,t)|\right)
< \infty,\tag3.47
$$
where $\gamma>0$ is a number which will be chosen later so that  that the
operator $V$ in (3.45) will be a contraction mapping on the Banach space of
functions with norm (3.47) for a fixed pair $A,B$. To choose $\gamma>0$, let us
estimate the norm of $V$. One has:
$$\multline
\Vert VL\Vert\leq\underset -\infty<\xi\leq A, 0\leq\eta\leq B\to\sup
\left(\int^\xi_{-\infty}ds\int^\eta_0dt|Q(s,t)|e^{-\gamma(\eta -t)}e^{-\gamma t}
|L(s,t)|\right)\leq \\
\leq\Vert L\Vert\underset -\infty <\xi\leq A,0\leq\eta
\leq B\to\sup\int^{\xi}_{-\infty}ds\int^\eta_0dt\left(2e^{s+t}+e^{s+t}
|q\left(
e^{\frac{s+t}2}\right)|\right)e^{-\gamma(\eta -t)}\leq\frac c\gamma\Vert L
\Vert,
\endmultline
\tag3.48
$$
where $c>0$ is a constant depending on $A,B$ and $\int^a_0r|q(r)|dr$.
Indeed, one has:
$$
2\int^A_{-\infty}ds\int^\eta_0dte^{s+t-\gamma(\eta-t)}=2e^A\int^\eta_0dte^{t-
\gamma (\eta -t)}dt\leq 2e^{A+B}\frac{1-e^{-\gamma B}}\gamma=\frac{c_1}\gamma,
\tag3.49'
$$
and, using the substitution $\sigma=e^{\frac{s+t}2}$, one gets:
$$
\multline
\int^A_{-\infty}ds\int^\eta_0dte^{s+t}|q(e^{\frac{s+t}2})|e^
{-\gamma(\eta -t)}=\\
=\int^\eta_0dte^{-\gamma (\eta -t)}\int^A_{-\infty}dse^{s+t}
\vert q\left(e^{\frac{s+t}2}\right)\vert=\\
=2\int^\eta_0dte^{-\gamma(\eta -t)}\int^{e^\frac{A+t}2}_0d\sigma\sigma|
q(\sigma )|=\\
=\frac{2(1-e^{-\gamma B})}\gamma\int^a_0d\sigma\sigma |q(\sigma)|:=
\frac{c_2}\gamma.
\endmultline
\tag3.49''
$$
From these estimates inequality (3.48) follows.

It follows from (3.48) that $V$ is a contraction mapping in the space
$X_{AB}$ of continuous functions in the region $-\infty<\xi\leq A,\quad
0\leq\eta\leq B$, with the norm (3.47) provided that
$$
\gamma>c.\tag3.50
$$
Therefore equation (3.45) has a unique solution $L(\xi,\eta)$ in the region
$$
-\infty<\xi<A,\qquad 0\leq\eta\leq B\tag3.51
$$
for any real $A$ and $B>0$
if (3.50) holds. This means that the above solution is defined for any
$\xi\in \Bbb R$ and any $\eta\geq 0$.

Equation (3.45) is equivalent to problem (3.41)-(3.43) and,
by (3.40), one has:
$$
B(\xi,\eta)=L(\xi,\eta)e^{\frac \xi 2}.\tag3.52
$$
Therefore we have proved the
existence and uniqueness of $B(\xi,\eta)$, that is, of
the kernel $\k =B(\xi,\eta )$ of the transformation operator (1.32). Recall
that $r$ and $\rho$ are related to $\xi$ and $\eta$ by formulas (3.31).

Let us formulate the result:
\proclaim{Lemma 3.2} The kernel of the transformation operator (1.32) solves
problem (3.27)-(3.29). The solution to this problem does exist and is unique
for any potential $q(r)\in C^1(0,a)$ in the class of twice continuously
differentiable functions. If $q(r)\in L^\infty (0,a)$, then $\k$ has first
derivatives which are bounded and equation (3.27) has to be understood in the
sense of distributions. The following estimate holds for any $r>0$:
$$
\iro|\k |\rho^{-1}d\rho<\infty.\tag3.53
$$
\endproclaim
\demo{Proof of Lemma 3.2} We have already proved all the assertions of Lemma
3.2 except estimate (3.53). Let us prove this estimate.

Note that
$$
\iro |\k |\rho^{-1}d\rho=
r\int^\infty_0|L(2\ln r-\eta,\eta)|e^{-\frac \eta 2}d\eta
<\infty\tag3.54
$$
Indeed, if $r>0$ is fixed, then, by (3.31), $\xi +\eta =2\ln r=const$. 
Therefore
$d\xi =-d\eta$, and $\rho^{-1}d\rho=
\frac 12(d\xi-d\eta)=-d\eta,\quad \xi =2\ln r-\eta$. Thus:
$$
\iro |\k|\rho^{-1}d\rho=
\int^\infty_0|L(2\ln r-\eta,\eta)|e^{\frac{2\ln r-\eta}2}d\eta =r
\int^\infty_0|L(2\ln r-\eta,\eta)|e^{-\frac \eta 2}d\eta.\tag3.55
$$
The following estimate holds:
$$
|L(\xi,\eta)|\leq c e^{(2+\epsilon_1) [\eta \mu_1 (\xi +
\eta)]^{\frac 12 +\epsilon_2}},\tag3.56
$$
where $\epsilon_j>0, \, j=1,2,$
are arbitrarily small numbers and $\mu_1$ is defined in
formula (3.60) below, see also formula (3.58) for the
definition of $\mu$.

Estimate (3.56) is proved  below, in Lemma
3.3.

From (3.55) and (3.56) estimate (3.53) follows. Lemma 3.2 is proved.$\square$
\enddemo
\proclaim{Lemma 3.3} Estimate (3.56) holds.
\endproclaim
\demo{Proof of Lemma 3.3} From (3.45) one gets:
$$
m(\xi,\eta)\leq c_0+(Wm)(\xi,\eta),\qquad
m(\xi,\eta):=|L(\xi,\eta)|,\tag3.57
$$
where $c_0=\underset -\infty<\xi<\infty\to \sup |b(\xi)|\leq\frac
12\int^a_0s
|q(s)|ds$ (see (3.42)), and
$$
Wm:=\int^\xi_{-\infty}ds\int^\eta_0dt\mu (s+t)m(s,t),
\quad \mu(s):=\frac 12 e^s\left
(1+|q(e^{\frac s2})|\right).\tag 3.58
$$
It is sufficient to consider inequality (3.57) with $c_0=1$: 
if $c_0=1$ and the
solution $m_0(\xi,\eta)$  to (3.57) satisfies (3.56)
with $c=c_1$, then the solution
$m(\xi,\eta)$ of (3.57) with any
$c_0>0$ satisfies (3.56) with $c=c_0c_1$.

Therefore, assume that $c_0=1$, then (3.57) reduces to:
$$
m(\xi,\eta)\leq1+(Wm)(\xi,\eta).\tag3.59
$$
Inequality (3.56) follows from (3.59) by iterations. Let us give the details.

Note that
$$
W1=\int^\xi_{-\infty}ds\int^\eta_0dt\mu (s+t)=\int^\eta_0dt\int^\xi_{-\infty}
ds\mu (s+t)=\int^\eta_0dt\mu_1(\xi +t)\leq\eta\mu_1(\xi+\eta).
$$
Here we have used the notation
$$
\mu_1(\xi)=\int^\xi_{-\infty}\mu (s) ds,\tag 3.60
$$
and the fact that $\mu_1(s)$ is a monotonically increasing function, since
$\mu
(s)>0$. Note also that $\mu_1(s)<\infty$ for any $s, \,-\infty<s<\infty$.

Furthermore,
$$
W^21\leq\int^\xi_{-\infty}ds\int^\eta_0dt\mu (s+t)t\mu_1(s+t)\leq\int^\eta_0dtt
\int^\xi_{-\infty}ds\mu (s+t)\mu_1(s+t)=\frac{\eta^2}{2!}\frac{\mu^2_1(\xi+\eta)}
{2!}.\tag3.61
$$
Let us prove by induction that
$$
W^n1\leq\frac{\eta^n}{n!}\frac{\mu^n_1(\xi +\eta)}{n!}.\tag3.62
$$
For $n=1$ and $n=2$ we have checked (3.62). Suppose (3.62) holds for some $n$,
then
$$
W^{n+1}1\leq W\left(\frac{\eta^n}{n!}\frac{\mu^n_1(\xi+\eta)}{n!}\right)=
\int^\eta_0dt\frac{t^n}{n!}\int^\xi_{-\infty}ds\mu (s+t)\frac{\mu^n_1(s+t)}{n!}
\leq\frac{\eta^{n+1}}{(n+1)!}\frac{\mu^{n+1}_1(\xi
+\eta)}{(n+1)!}.\tag3.63
$$
By induction, estimate (3.61) is proved for all $n=1,2,3,...$. Therefore (3.59)
implies
$$
m(\xi,\eta)\leq 1+\sum^\infty_{n=1}\frac{\eta^n}{n!}\frac{\mu^n_1(\xi +\eta)}
{n!}\leq c e^{(2+\epsilon_1) [\eta \mu_1 (\eta+\xi)]^{\frac 12
+\epsilon_2}}, \tag3.64
$$
where we have used Theorem 2 from [L, section 1.2], namely
the order of the entire function $F(z):=1+\sum_{n=1}^\infty \frac
{z^n}{(n!)^2}$ is $\frac 12$ and its type is 2. The constant
$c>0$ in (3.56) depends on $\epsilon_j, \, j=1,2.$

Recall that the order of an entire function $F(z)$ is the number
$\rho:=limsup_{r\to \infty}\frac {ln\,ln\,M_F(r)}{ln\,r}$,
where $M_F(r):=max_{|z|=r}|F(z)|$. The type of $F(z)$ is the number
$\sigma:=limsup_{r\to \infty}\frac {ln\, M_F(r)}{r^\rho}$.
It is known [L], that if $F(z)=\sum_{n=0}^\infty c_nz^n$ is
an entire function, then its order $\rho$ and type $\sigma$
can be calculated by the formulas:
$$\rho=limsup _{n\to \infty} \frac {n\, ln\,n}{ln\,\frac 1{|c_n|}},
\quad \sigma= 
\frac { limsup_{n\to \infty}( n|c_n|^{\frac {\rho} {n}})}{e\rho}.
$$
If $c_n=\frac 1 {(n!)^2}$, then the above formulas yield $\rho=\frac 12$
and $\sigma=2$.
Lemma 3.3 is proved. $\qed$
\enddemo
\subhead 3.5. Proof of Theorem 2.1\endsubhead

Suppose that there are two potentials which generate the same data $\dl_
{\forall\ell\in\Cal L}$. In section 3.1 we have proved that this implies (3.3).
From (3.3) and (1.26) it follows that (3.3) is equivalent to (1.28).

From Lemma 3.2, formula (1.32), Lemma 1.1, and the definition
$$
h(\ell)=\int^a_0drp(r)\varphi_{1\ell}(r)\varphi_{2\ell}(r),\tag3.65
$$
it follows that the function 
$$h_1(\ell):=
\bigl [\sqrt{\frac 2\pi} \Gamma\left(\frac 12\right)
\Gamma (\ell +1)2^{\ell +\frac 12}\bigr ]^2 h(\ell)\in N.  
\tag3.65' $$ 
This is
checked as in the proof of Lemma 1.1 in section 3.2. There are
four terms which one gets from multiplication of 
$\varphi_{1\ell}(r)$ by $\varphi_{2\ell}(r)$, where
$\varphi_{j\ell}(r),\, j=1,2,$ are expressed by formula
(1.32) with $K(r, \rho)= K_j(r, \rho),\, j=1,2.$
The first term contains $u_{\ell}^2(r)$ and is identical
with (1.30), the second and third terms contain the products
of the type $u_{\ell}(r) u_{\ell}(\rho)$,
while the fourth term contains the term
$u_{\ell}(\rho_1) u_{\ell}(\rho_2)$.  These terms are treated
like in the proof of Lemma 1.1. and estimate (3.53)
is used.

From (3.65'),  Theorem 1.3, and 
assumption (1.28) it follows that
$h(\ell)=0$ for $\ell=0,1,2,3,\dots$.

From this and (1.26) it follows that (1.23) holds for $\ell=
0,1,2,3, \dots$.

From (1.23) and (1.22) it follows that
$$
\int_{B_a}dxp(r)\psi_1(x,\alpha)\overline {\psi_2(x,\beta)}=
0
\quad\forall\alpha,\beta\in
S^2.\tag3.66
$$
From (3.66) and Theorem 1.2 one concludes that $p(r)=0$.

Theorem 2.1 is proved. $\qed$
 
\subhead 3.6. Heuristic motivation of the basic result\endsubhead

Here we give an heuristic motivation of 
the basic result, namely of Theorem 2.1.

It is well known that
$$
\ul (r):=\sqrt{\frac{\pi r}2}J_{\ell+\frac 12}(r)=\sqrt{\frac r2}\left(\frac{er}
{2\ell +1}\right)^{\frac{2\ell +1}2}\frac 1{\sqrt{2\ell +1}}[1+o(1)]\text{ as }
\ell\rightarrow\infty.\tag3.67
$$
One can prove that $\varphi_{\ell}(r)$ has the same asymptotics
$$
\varphi_\ell (r)=\sqrt{\frac r2}\left(\frac{er}{2\ell +1}\right)^{\frac{2\ell
+1}2}\frac 1{\sqrt{2\ell +1}}[1+o(1)]\text{ as }\ell\rightarrow\infty.\tag3.68
$$
If one substitutes (3.68) into (1.28), one gets
$$
0=\int^a_0drr^2p(r)r^{2\ell}[1+o(1)]\quad\forall\ell\in\Cal L.\tag3.69
$$
If one neglects the term $o(1)$, then one gets
$$
0=\int^a_0drr^2p(r)r^{2\ell}\quad\forall\ell\in\Cal L.\tag3.70
$$
From (3.70) and the well known M\"untz's theorem [Ru, p.336], it follows that
$p(r)=0$, which yields the conclusion of Theorem 2.1.

This heuristic argument is not a proof because the justification of the
passage
from (3.69) to (3.70) is not given and is not clear if such a justification
 can
be given directly. Our proof of Theorem 2.1 can be considered as an indirect
justification of this heuristic argument.

It is known that condition (1.1) is necessary and sufficient for completeness
of the set $\{r^\ell\}_{\ell\in\Cal L}$ in $L_1(0,a)$ for any fixed $a>0$.

Therefore one can raise the following question: 

{\it Is it true that condition (1.1)
is necessary for the conclusion of Theorem 2.1 to be valid?} 

This interesting
question is open.

\vfill

\enddocument